\documentclass{jfm}
\usepackage[utf8]{inputenc}

\renewcommand{\pagelimitfooter}{}
\renewcommand{\absfooterflag}{}

\newcounter{ncorresp}
\let\Oldcorresp\corresp
\renewcommand{\corresp}[1]{{\Oldcorresp{#1}}\stepcounter{ncorresp}}

\usepackage{graphicx}
\usepackage{newtxtext}
\usepackage{newtxmath}
\usepackage{natbib}
\usepackage{hyperref}
\hypersetup{
    colorlinks = true,
    urlcolor   = blue,
    citecolor  = black,
}

\newcommand{\RomanNumeralCaps}[1]
\linenumbers

\usepackage{aas_macros} 
\usepackage{mathtools} 
\usepackage{siunitx}

\usepackage[inline]{enumitem} 

\renewcommand{\dh}{\partial}
\renewcommand{\d}{\mathrm{d}}
\newcommand{\dive}{\nabla\!\cdot}
\newcommand{\curl}{\nabla\!\times\!}
\newcommand{\grad}{\nabla}

\newcommand{\defn}{\equiv}
\newcommand{\abs}[1]{\left|#1\right|}

\newcommand{\bigO}{O} 
\newcommand{\re}{\operatorname{Re}}

\renewcommand{\vec}[1]{\boldsymbol{#1}}
\newcommand{\cross}{\times}
\newcommand{\Dirac}{\operatorname{\delta}}
\newcommand{\continuedTerm}{\phantom{===}}
\newcommand{\negphantom}[1]{\settowidth{\dimen0}{#1}\hspace*{-\dimen0}} 

\newcommand{\emf}{\mathcal{E}}
\newcommand{\mesoAvBr}[1]{\left<#1\right>}
\newcommand{\mesoAv}[1]{\overline{#1}}

\newcommand{\FT}[1]{\widetilde{#1}} 
\newcommand{\dynOne}{\mathcal{D}_1}
\newcommand{\dynTwo}{\mathcal{D}_2}
\newcommand{\evCor}{g}
\newcommand{\odCor}{h}
\newcommand{\arOne}{\chi_1}
\newcommand{\arTwo}{\chi_2}
\newcommand{\dynOneA}{\widetilde{\mathcal{D}}_1}

\title{Mean-field dynamo due to spatiotemporal fluctuations of the turbulent kinetic energy}
\author{
Kishore Gopalakrishnan
\corresp{\email{kishoreg@iucaa.in}}
\and
Nishant K. Singh
\corresp{\email{nishant@iucaa.in}}
}
\affiliation{Inter-University Centre for Astronomy \& Astrophysics, Post Bag 4, Ganeshkhind, Pune 411 007, India}

\begin{document}

\maketitle
\addtocounter{footnote}{\value{ncorresp}}

\begin{abstract}
	In systems where the standard $\alpha$ effect is inoperative, one often explains the existence of mean magnetic fields by invoking the `incoherent $\alpha$ effect', which appeals to fluctuations of the mean kinetic helicity at a mesoscale.
	Most previous studies, while considering fluctuations in the mean kinetic helicity, treated the mean turbulent kinetic energy at the mesoscale as a constant, despite the fact that both these quantities involve second-order velocity correlations.
	The mean turbulent kinetic energy affects the mean magnetic field through both turbulent diffusion and turbulent diamagnetism.
	In this work, we use a double-averaging procedure to analytically show that fluctuations of the mean turbulent kinetic energy at the mesoscale (giving rise to $\eta$-fluctuations at the mesoscale, where the scalar $\eta$ is the turbulent diffusivity) can lead to the growth of a large-scale magnetic field even when the kinetic helicity is zero pointwise.
	Constraints on the operation of such a dynamo are expressed in terms of dynamo numbers that depend on the correlation length, correlation time, and strength of these fluctuations.
	In the white-noise limit, we find that these fluctuations reduce the overall turbulent diffusion, while also contributing
	a drift term which does not affect the growth of the field.
	We also study the effects of nonzero correlation time and anisotropy.
	Turbulent diamagnetism, which arises due to inhomogeneities in the turbulent kinetic energy, leads to growing mean field solutions even when the $\eta$-fluctuations are statistically isotropic.
\end{abstract}



\section{Introduction}
Astrophysical magnetic fields are observed on galactic, stellar, and planetary scales \citep{KanduPhysicsReports2005, jones11}.
Some stars even exhibit periodic magnetic cycles.
The Earth itself has a dipolar magnetic field that shields it from the solar wind.
Dynamo theory studies the mechanisms behind the generation and maintenance of these large-scale magnetic fields by fluid flows at much smaller scales \citep{RuzShuSok88, KanduPhysicsReports2005, jones11, rincon19}.
Mean-field magnetohydrodynamics takes advantage of scale-separation to make the problem analytically tractable \citep{MoffatMagFieldGenBook, KrauseRadler80}.

The turbulent electromotive force, which is determined by correlations between the fluctuating velocity and magnetic fields, plays a crucial role in mean-field dynamo theory.
For homogeneous and isotropic turbulence, using the quasilinear approximation, one can express the turbulent electromotive force in terms of the turbulent transport coefficients $\alpha$ (which is proportional to the mean kinetic helicity) and $\eta$ (the turbulent diffusivity, which is proportional to the mean kinetic energy) when the magnetic field is weak \citep[chapter 7]{MoffatMagFieldGenBook}.
The contribution of $\alpha$, if nonzero, may cause growth of the mean magnetic field, while $\eta$ always dissipates it when the turbulence is homogeneous.

Even when the mean kinetic helicity is zero, \citet{kraichnan1976diffusion} found that fluctuations of the kinetic helicity can suppress the turbulent diffusivity.
If the fluctuations are strong or long-lived enough, the effective diffusivity may become negative, leading to growth of the large-scale magnetic field (\citealp{kraichnan1976diffusion}; \citealp[sec.~7.11]{MoffatMagFieldGenBook}; \citealp{singh2016moffatt}).
This effect, usually referred to as the `incoherent $\alpha$ effect', has also been studied in combination with shear \citep{sokolov97, vishniac97, silantev2000, sridhar14}.
The `incoherent $\alpha$-shear dynamo' has been invoked \citep{brandenburg2008alphaeta} to explain the generation of a large-scale magnetic field in simulations of nonhelical turbulence with background shear \citep{yousef08, SinJin15}.

To derive his result, \citet{kraichnan1976diffusion} used a process of double-averaging, where one first obtains the mean-field equations at some mesoscale, and then fluctuations of the mesoscale transport coefficients may lead to effects at some larger scale upon subsequent averaging.
There are two viewpoints (not mutually exclusive) on the applicability of this method.
One is that we require the system to have scale separation, such that the turbulent spectra peak at some small scale, while averaged quantities themselves fluctuate at some mesoscale, and then there exists an even larger scale where a magnetic field can grow \citep[e.g.][p.~178]{MoffatMagFieldGenBook}.
As an example of a physical system where such a picture may be relevant, we note that in the solar photosphere, mesoscales can be identified with granulation or supergranulation \citep[for a review on supergranulation, see][]{RinRie18}.
Another viewpoint is to think of multiscale averaging as a renormalization procedure which tells us something about the contributions of higher moments of the velocity field to the turbulent transport coefficients (\citealp[e.g.][sec.~11]{Moffatt83}; \citealp[p.~341]{silantev2000}).
In support of this, we note that \citet{knobloch77}\footnote{
\Citet{nicklaus88} point out some errors in this paper.
} and \citet{nicklaus88} have used a cumulant expansion to calculate the lowest-order corrections to the quasilinear approximation.
In agreement with the results obtained by multiscale averaging, they find that the turbulent diffusivity is suppressed.

Regardless of one's viewpoint, it seems natural to wonder why fluctuations of the helicity should have a more privileged position than fluctuations of the kinetic energy
(i.e.\@ the turbulent magnetic diffusivity).
Dynamos can be driven or boosted by spatial variations of the \emph{microscopic} magnetic diffusivity or the magnetic permeability \citep{BusWic92, RogMcE17, GieNorSte10, PetAleGis16, GreRudEls23}.
Further, in simulations, it is found that fluctuations of $\alpha$ coexist with fluctuations of $\eta$ \citep[e.g.][fig.~10]{brandenburg2008alphaeta}.
While \citet{silantev99,silantev2000} has considered fluctuations of the turbulent diffusivity (and found that the effective turbulent diffusivity is suppressed), he has not included the effect of turbulent diamagnetism (expulsion of the magnetic field from turbulent regions); the latter is a natural consequence of spatial variations of the turbulent kinetic energy, and thus cannot be ignored.

Here, we explore the effects of mesoscale fluctuations of the turbulent magnetic diffusivity, with nonzero correlation time, on the evolution of the large-scale magnetic field.
The procedure we follow is the same as that of \citet{singh2016moffatt}.

In section \ref{section: general derivation}, we derive the evolution equation for the large-scale magnetic field, along with an expression for its growth rate, by using the quasilinear approximation.
In section \ref{section: dyn nums isotrop fluc}, we simplify the expression for the growth rate, assuming the fluctuations of $\eta$ are isotropic.
In section \ref{B.etaFluc: section: anisotropy}, we explain how the growth rate is modified by anisotropy.
In section \ref{section: general suppression turb diff}, we relate the growth in some regimes to a negative effective turbulent diffusivity.
In section \ref{B.etaFluc: estimate dynamo numbers}, we show how to estimate the dynamo numbers in astrophysical systems, taking the solar photosphere as an example.
Finally, we discuss the implications of our results and possible future directions in section \ref{section: conclusions}.

\section{Derivation of the evolution equation and the growth rate}
\label{section: general derivation}

\subsection{Setup and assumptions}
The mean magnetic field, $\vec{B}$, evolves according to \citep[e.g.][eq.~7.7]{MoffatMagFieldGenBook}
\begin{equation}
	\frac{\dh\vec{B} }{\dh t} = \curl\left( \vec{V}\cross\vec{B} + \vec{\emf} \right) + \eta_m \nabla^2 \vec{B}
	\label{B.etaFluc: eq: mean field induction general}
\end{equation}
where $\vec{V}$ is the mean velocity; $\eta_m$ is the microscopic magnetic diffusivity; and $\vec{\emf}$, the turbulent electromotive force (EMF), is related to the correlation between the fluctuating velocity and magnetic fields.

Since the MHD equations are nonlinear, the evolution equations for moments of a particular order depend on moments of higher orders.
For example, the EMF in equation \ref{B.etaFluc: eq: mean field induction general} is a double-correlation of the fluctuating fields.
To keep the system of equations manageable, one has to truncate this hierarchy by applying a \emph{closure}.
To avoid solving for the fluctuating magnetic field, one requires an expression for the EMF in terms of the mean magnetic field itself.
If the mean magnetic field is weak and varies slowly, one typically assumes that the EMF depends only on the mean magnetic field and its first derivatives, obtaining a general expression of the form $\emf_i = \alpha_{ij} B_j + \eta_{ijk} \dh_j B_k$.
Here, and in what follows, repeated indices are summed over.
The tensors $\alpha_{ij}$ and $\eta_{ijk}$ may depend on the statistical properties of the velocity field.
The expressions for these tensors depend on the closure used.

One of the most widely used closures in dynamo theory is the quasilinear approximation (also called the first-order smoothing approximation, FOSA; or the second-order correlation approximation, SOCA) (e.g.\@ \citealp[sec.~7.5]{MoffatMagFieldGenBook}; \citealp[sec.~4.3]{KrauseRadler80}).
The quasilinear approximation is rigorously valid only when either the magnetic Reynolds number (the ratio of the diffusive to the advective timescale) or the Strouhal number (the ratio of the velocity correlation time to its turnover time) are small \citep[p.~49]{KrauseRadler80}.
The former is never small in the astrophysical systems of interest, while it is unclear if the latter is small.
Nevertheless, in the context of mean-field dynamo theory, the quasilinear approximation often remains qualitatively correct well outside its domain of formal validity.
More complicated closures such as the EDQNM closure \citep[e.g.][]{pouquet76} and the DIA \citep{kraichnan1977} are extremely difficult to work with.

For weakly inhomogeneous nonhelical turbulence, the EMF is given in the quasilinear approximation by \citep[eq.~3.11]{robertsSoward75}
\begin{equation}
	\vec{\emf} = - \frac{1}{2} \nabla \eta \cross \vec{B} - \eta \curl\vec{B} \label{B: eq: diamagnetic pumping}
\end{equation}
where $\eta$ is the turbulent diffusivity (proportional to the turbulent kinetic energy).
We note that \citet{silantev99,silantev2000} did not consider the first term above.\footnote{
Appendix \ref{appendix: no pumping} describes how the absence of this term qualitatively changes the behaviour of the system.
}
Comparing this term with the $\vec{V}\cross\vec{B}$ term in equation \ref{B.etaFluc: eq: mean field induction general}, we see that the former can be thought of as describing an effective velocity $-\grad\eta/2$ that acts on the mean magnetic field.
This transports the magnetic field in the direction in which the turbulent kinetic energy decreases.
By analogy with the reduction of the magnetic field in diamagnetic materials, this is usually referred to as `turbulent diamagnetism' or `diamagnetic pumping'.
For stratified turbulence, or in the presence of small-scale magnetic fields, additional terms arise \citep{vainshtein1983macroscopic}, but we ignore those effects in this work.
As mentioned in the introduction, the effect of helical turbulence ($\alpha$) has been extensively studied, so we restrict ourselves to turbulence that is nonhelical pointwise.
The first term of equation \ref{B: eq: diamagnetic pumping} may be considered one of many contributions to the off-diagonal components of $\alpha_{ij}$.
In the literature, such contributions are sometimes described in terms of a vector $\vec{\gamma}$; see \citet[eq.~42]{radler2003}.
While the calculations we present can be carried out using the $\alpha_{ij}$ and $\eta_{ijk}$ tensors in their full glory, we use a simpler expression in order to keep the results interpretable.

Although the mean-field approach does not formally require scale-separation, we associate averages with length/time scales for clarity of exposition.
Let us assume that $\eta$ fluctuates at length/time scales (henceforth referred to as the mesoscales) much larger than the scales at which the turbulent velocity fluctuates.
We employ a double-averaging approach \citep{kraichnan1976diffusion, singh2016moffatt}, in which we treat $\eta$ (at the mesoscale) as a stochastic scalar field which is a function of both position and time (i.e. $\eta = \eta(\vec{x},t)$).
For any mesoscale quantity $\Box$, we use $\mesoAvBr{\Box}$ and $\mesoAv{\Box}$ to denote its averages at the larger scale.
We assume this average satisfies Reynolds' rules \citep[e.g.][sec.~3.1]{MoninYaglomVol1}.

If we set the mean velocity to zero, ignore the microscopic diffusivity (which is usually much smaller than the turbulent diffusivity in the systems of interest), and use equation \ref{B: eq: diamagnetic pumping}, we can write equation \ref{B.etaFluc: eq: mean field induction general}, the evolution equation for the mesoscale magnetic field, as
\begin{equation}
	\frac{\dh\vec{B}}{\dh t} = \curl\left( - \frac{1}{2} \nabla \eta \cross \vec{B} - \eta \curl\vec{B} \right)
	\label{B.etaFluc: eq: evolution meso B}
\end{equation}

In section \ref{section: assume homo sep}, we assume the fluctuations of $\eta$ are statistically homogeneous, stationary, and separable in order to obtain an integro-differential equation for the large-scale magnetic field.
In section \ref{section: evoleq O(tau^1)}, we simplify this equation by assuming the fluctuations of $\eta$ are white noise, while in section \ref{section: evoleq O(tau^2)}, we also keep terms linear in the correlation time of $\eta$.

\subsection{Evolution equation in Fourier space}
We now move to Fourier space with
\begin{align}
	\FT{f}(\vec{k},t) \defn \int \frac{\d\vec{x}}{(2\pi)^3}\, e^{i\vec{k}\cdot\vec{x}} f(\vec{x},t) \label{B: eq: Fourier transform definition eta fluctuations}
\end{align}
where we have used a tilde to denote the spatial Fourier transform of a quantity.
The convolution theorem takes the form
\begin{equation}
	\int\frac{\d\vec{x}}{(2\pi)^3}\, e^{i\vec{k}\cdot\vec{x}} f(\vec{x}) g(\vec{x}) = \int\d\vec{p} \, \FT{f}(\vec{p}) \FT{g}(\vec{k}-\vec{p})
\end{equation}
Equation \ref{B.etaFluc: eq: evolution meso B} then becomes (omitting the temporal arguments whenever there is no ambiguity)
\begin{align}
	\frac{\dh \FT{\vec{B}}(\vec{k}) }{\dh t}
	={}&
	\vec{k} \cross \int\d\vec{p} \left(\frac{\vec{k}+\vec{p}}{2}\right) \cross \FT{\vec{B}}(\vec{p}) \, \FT{\eta}(\vec{k}-\vec{p})
\end{align}
Taking the average of the above, we obtain
\begin{align}
	\frac{\dh}{\dh t} \mesoAvBr{ \FT{\vec{B}}(\vec{k}) }
	={}&
	\vec{k} \cross \int\d\vec{p} \left(\frac{\vec{k}+\vec{p}}{2}\right) \cross \bigg( \mesoAvBr{ \FT{\eta}(\vec{k}-\vec{p}) }\mesoAvBr{ \FT{\vec{B}}(\vec{p}) } + \mesoAvBr{ \FT{\mu}(\vec{k}-\vec{p}) \,\FT{\vec{b}}(\vec{p}) } \bigg)
	\label{B: eq: mean B multiscale varying eta}
\end{align}
where we have split the mesoscale fields into their mean and fluctuating parts, i.e.\@ $\FT{\vec{B}} = \mesoAvBr{\FT{\vec{B}}} + \FT{\vec{b}}$ and $\FT{\eta} = \mesoAvBr{\FT{\eta}} + \FT{\mu}$.
We write the equation for $\FT{\vec{b}}$ as
\begin{align}
	\begin{split}
		\frac{\dh \FT{\vec{b}}(\vec{k}) }{\dh t} = 
		\vec{k} \cross \int\d\vec{p} \left(\frac{\vec{k}+\vec{p}}{2}\right) \cross \bigg( 
		& \mesoAvBr{ \FT{\eta}(\vec{k}-\vec{p}) } \FT{\vec{b}}(\vec{p})
		+ \FT{\mu}(\vec{k}-\vec{p}) \mesoAvBr{ \FT{\vec{B}}(\vec{p}) }
		\\& + \FT{\mu}(\vec{k}-\vec{p}) \,\FT{\vec{b}}(\vec{p})
		- \mesoAvBr{ \FT{\mu}(\vec{k}-\vec{p}) \,\FT{\vec{b}}(\vec{p}) }
		\bigg)
	\end{split}
\end{align}
We now apply the quasilinear approximation, where the equations for the fluctuating fields are truncated by keeping only terms which are at most linear in the fluctuating fields.
We then obtain
\begin{align}
	\begin{split}
		\frac{\dh \FT{\vec{b}}(\vec{k}) }{\dh t} ={}&
		\vec{k} \cross \int\d\vec{p} \left(\frac{\vec{k}+\vec{p}}{2}\right) \cross \bigg( 
		\mesoAvBr{ \FT{\eta}(\vec{k}-\vec{p}) } \FT{\vec{b}}(\vec{p})
		+ \FT{\mu}(\vec{k}-\vec{p}) \mesoAvBr{ \FT{\vec{B}}(\vec{p}) }
		\bigg)
	\end{split} \label{B: eq: dbBYdt fourier space}
\end{align}

\subsection{Homogeneity and separability}
\label{section: assume homo sep}
To simplify the preceding expression, we assume that the moments of $\eta(\vec{x},t)$ are statistically homogeneous and stationary.
We can then write, say, $\mesoAvBr{\mu(\vec{x},\tau_1) \mu(\vec{y},\tau_2)} = C(\vec{x} - \vec{y}, \tau_1 - \tau_2)$.
Further, we assume that $\mesoAvBr{\mu(\vec{x},\tau_1) \mu(\vec{y},\tau_2)} $ can be written as the product of a temporal correlation function and a spatial correlation function, i.e.\@ $C(\vec{x} - \vec{y}, \tau_1 - \tau_2) = Q(\vec{x} - \vec{y}) \, S(\tau_1 - \tau_2)$.
In Fourier space, these assumptions can be expressed as
\begin{subequations}
	\begin{align}
		\mesoAvBr{\FT{\eta}(\vec{k},t)} ={}& \mesoAv{\eta} \Dirac(\vec{k})
		\\
		\mesoAvBr{\FT{\mu}(\vec{p},\tau_1) \FT{\mu}(\vec{q},\tau_2)} ={}& \FT{Q}\!\left(\vec{p}\right) S(\tau_1-\tau_2) \Dirac(\vec{p}+\vec{q})
		\label{B: eq: homogeneous separable mumu at large scale, fourier space}
	\end{align}
	\label{B: eq: homogeneous eta at large scale, fourier space}%
\end{subequations}
For $S$, we require
\begin{equation}
	2 \int_0^\infty S(t) \, \d t = 1
\end{equation}
and define the correlation time of $\eta$ as
\begin{equation}
	\tau_\eta \defn 2 \int_0^\infty t S(t) \, \d t
\end{equation}
We can then write
\begin{align}
	\begin{split}
		\vec{k} \cross \int\d\vec{p} \left(\frac{\vec{k}+\vec{p}}{2}\right) \cross \FT{\vec{b}}(\vec{p}) \mesoAvBr{ \FT{\eta}(\vec{k}-\vec{p}) } 
		=
		- \mesoAv{\eta} k^2 \FT{\vec{b}}(\vec{k})
	\end{split} \label{B.etaFluc: eq: simp cross cross integral homo mean eta}
\end{align}
and
\begin{equation}
	\vec{k} \cross \int\d\vec{p} \left(\frac{\vec{k}+\vec{p}}{2}\right) \cross \mesoAvBr{ \FT{\vec{B}}(\vec{p}) } \mesoAvBr{ \FT{\eta}(\vec{k}-\vec{p}) }
	=
	- \mesoAv{\eta} k^2 \mesoAvBr{ \FT{\vec{B}}(\vec{k}) }
	\label{B.etaFluc: eq: simp cross cross integral meanB homo mean eta}
\end{equation}

Using equation \ref{B.etaFluc: eq: simp cross cross integral homo mean eta}, equation \ref{B: eq: dbBYdt fourier space} can be written as
\begin{align}
	\begin{split}
		\frac{\dh \FT{\vec{b}}(\vec{k}) }{\dh t}
		={}& 
		- \mesoAv{\eta} k^2 \FT{\vec{b}}(\vec{k})
		+ \vec{k} \cross \int\d\vec{p} \left(\frac{\vec{k}+\vec{p}}{2}\right) \cross \mesoAvBr{ \FT{\vec{B}}(\vec{p}) }
		\FT{\mu}(\vec{k}-\vec{p})
	\end{split}
\end{align}
which gives us
\begin{align}
	\begin{split}
		\FT{\vec{b}}(\vec{k},t)
		={}&
		\vec{k} \cross \int_0^t \d\tau \int\d\vec{p} \, e^{- \mesoAv{\eta} k^2 \left(t-\tau \right)}
		\left( \frac{\vec{k} + \vec{p}}{2} \right) \cross \mesoAvBr{ \FT{\vec{B}}(\vec{p}, \tau) } \FT{\mu}(\vec{k}-\vec{p}, \tau)
		\\& +
		\FT{\vec{b}}(\vec{k},0)
	\end{split}
\end{align}
We assume that the initial fluctuations of the mesoscale magnetic field are uncorrelated with $\mu$.
Using the above along with equation \ref{B: eq: homogeneous separable mumu at large scale, fourier space}, we can write
\begin{align}
	\begin{split}
		\mesoAvBr{ \FT{\mu}(\vec{q},t) \, \FT{\vec{b}}(\vec{k},t) }
		=
		\vec{k} \cross \int_0^t \d\tau \,
		e^{- \mesoAv{\eta} k^2 \left(t-\tau \right)}  
		\left( \vec{k} + \frac{\vec{q}}{2} \right) \cross \mesoAvBr{ \FT{\vec{B}}(\vec{k}+\vec{q}, \tau) }
		\FT{Q}\!\left(\vec{q}\right) S(t-\tau) 
	\end{split} \label{B: eq: eta fluctuations multiscale average general expression mu b correlator}
\end{align}
Putting the above in equation \ref{B: eq: mean B multiscale varying eta} gives us an equation for $\mesoAvBr{ \FT{\vec{B}}(\vec{k}+\vec{q}, \tau) }$.
However, this is an integro-differential equation which is difficult to solve in general.
The resulting equation can be simplified by assuming $\tau_\eta$ is small.
In section \ref{section: evoleq O(tau^1)}, we assume $\tau_\eta=0$ and simplify the evolution equation for the large-scale magnetic field.
In section \ref{section: evoleq O(tau^2)}, we simplify the evolution equation neglecting $\bigO(\tau_\eta^2)$ terms.

\subsection{Evolution equation with white-noise fluctuations}
\label{section: evoleq O(tau^1)}
Assuming $S(t) = \Dirac(t)$, we write equation \ref{B: eq: eta fluctuations multiscale average general expression mu b correlator} as
\begin{equation}
		\mesoAvBr{ \FT{\mu}(\vec{q},t) \, \FT{\vec{b}}(\vec{k},t) }
		=
		\frac{1}{2}
		\, \vec{k} \cross \left[ \left( \vec{k} + \frac{\vec{q}}{2} \right) \cross \mesoAvBr{ \FT{\vec{B}}(\vec{k}+\vec{q}, t) } \right] 
		\FT{Q}\!\left(\vec{q}\right)
\end{equation}
Recalling that $\vec{k}\cdot\mesoAvBr{\FT{\vec{B}}(\vec{k}, t)} = 0$, we can use the above to write
\begin{align}
	\begin{split}
		\continuedTerm&\negphantom{\continuedTerm}
		\vec{k} \cross \int \d\vec{p} \left(\frac{\vec{k}+\vec{p}}{2}\right)\cross \mesoAvBr{ \FT{\mu}(\vec{k}-\vec{p},t) \, \FT{\vec{b}}(\vec{p},t) }
		\\
		={}& \int\d\vec{s} \, \frac{1}{8} \, \FT{Q}\!\left(\vec{s}\right)  \left( 4 k^4  - 8k^2 \vec{k}\cdot\vec{s} + 3 \left( \vec{k}\cdot\vec{s} \right)^2 + 2 k^2 s^2 - s^2 \vec{k}\cdot\vec{s} \right) \mesoAvBr{ \FT{\vec{B}}(\vec{k},t) }
	\end{split}
\end{align}
where $\vec{s} \defn \vec{k}-\vec{p}$.
Defining
\begin{equation}
	A^{(0)} \defn Q(\vec{0})
	\,,\,
	A^{(1)}_i \defn \left. \frac{\dh Q(\vec{\xi}) }{\dh \xi_i} \right|_{\vec{\xi} = \vec{0}}
	\,,\,
	A^{(2)}_{ij} \defn \left. \frac{\dh^2 Q(\vec{\xi}) }{\dh \xi_i \dh \xi_j} \right|_{\vec{\xi} = \vec{0}}
	\,,\,
	A^{(3)}_{i} \defn \left. \frac{\dh^3 Q(\vec{\xi}) }{\dh \xi_i \dh \xi_j \dh \xi_j} \right|_{\vec{\xi} = \vec{0}}
	\label{B.etaFluc: A defn}
\end{equation}
we write
\begin{align}
	\begin{split}
		\continuedTerm&\negphantom{\continuedTerm}
		\vec{k} \cross \int \d\vec{p} \left(\frac{\vec{k}+\vec{p}}{2}\right)\cross \mesoAvBr{ \FT{\mu}(\vec{k}-\vec{p},t) \, \FT{\vec{b}}(\vec{p},t) }
		\\
		={}& \frac{1}{8} \left( 4 A^{(0)} k^4  - 8 i A^{(1)}_i k^2 k_i - 3 A^{(2)}_{ij} k_ik_j - 2 A^{(2)}_{ii} k^2  + i A^{(3)}_i k_i \right) \mesoAvBr{ \FT{\vec{B}}(\vec{k},t) }
	\end{split}
\end{align}
Note that the functions defined in equation \ref{B.etaFluc: A defn} depend only on the value of $Q(\vec{\xi})$ and its spatial derivatives at the origin.
Putting this in equation \ref{B: eq: mean B multiscale varying eta} and using equation \ref{B.etaFluc: eq: simp cross cross integral meanB homo mean eta}, we obtain
\begin{align}
	\begin{split}
		\frac{\dh}{\dh t} \mesoAvBr{ \FT{\vec{B}}(\vec{k}) } 
		={}&
		\left( - \mesoAv{\eta} k^2 + \evCor(\vec{k}) \right) \mesoAvBr{ \FT{\vec{B}}(\vec{k}) }
		+ i \odCor(\vec{k}) \mesoAvBr{ \FT{\vec{B}}(\vec{k}) }
	\end{split} \label{B: eq: fourier space evolution B multiscale final tau_n zero}
\end{align}
where
\begin{subequations}
	\begin{align}
		\begin{split}
			\evCor(\vec{k})
			\defn{}&
			- \frac{3}{8} A^{(2)}_{ij} k_i k_j - \frac{1}{4} A^{(2)}_{ii} k^2 + \frac{1}{2} A^{(0)} k^4
		\end{split}
		\\
		\begin{split}
			\odCor(\vec{k})
			\defn{}&
			\frac{1}{8} A^{(3)}_i k_i - A^{(1)}_i k^2 k_i
		\end{split}
	\end{align} \label{eq: evCor OdCor definition}%
\end{subequations}
We see that $\evCor(\vec{k})$ describes corrections to the turbulent diffusivity (along with a $k^4$ hyperdiffusive term), while the term involving $\odCor(\vec{k})$ describes advection of the large-scale magnetic field with an effective velocity $A^{(3)}_i /8 - A^{(1)}_i k^2$.

To aid the interpretation of equation \ref{B: eq: fourier space evolution B multiscale final tau_n zero}, we note that if the spatial correlation function of the fluctuations of $\eta$ is an isotropic Gaussian (see appendix \ref{appendix: example gauss correlation function}), we can write
\begin{equation}
	\frac{\dh}{\dh t} \mesoAvBr{ \FT{\vec{B}}(\vec{k}) } 
	=
	- \left( \mesoAv{\eta} - \frac{9}{8} \, \beta \, \right) k^2 \mesoAvBr{ \FT{\vec{B}}(\vec{k}) }
	+ \bigO(k^4)\;,
	\label{B:.etaFluc: white noise evol assume simple corrfun}
\end{equation}
where $\beta \defn A^{(0)}/l_c^2 > 0$ represents the diffusivity arising from fluctuations of $\eta$ with a correlation length $l_c$. Thus, we find that fluctuations of $\eta$ reduce the turbulent diffusion of the large-scale magnetic field.

\subsection{Evolution equation with nonzero correlation time}
\label{section: evoleq O(tau^2)}
We expand
\begin{align}
	\mesoAvBr{ \FT{\vec{B}}(\vec{p}, \tau) } = \mesoAvBr{ \FT{\vec{B}}(\vec{p}, t) } - \left( t - \tau \right) \frac{ \dh }{\dh t}\!\mesoAvBr{ \FT{\vec{B}}(\vec{p}, t) } + \bigO( (t-\tau)^2 )
\end{align}
The idea is that when we substitute this into equation \ref{B: eq: eta fluctuations multiscale average general expression mu b correlator}, assume $t\gg \tau_\eta$, and perform the time integral, the powers of $(t-\tau)$ become powers of $\tau_\eta$.
The convergence of this series requires that the large-scale magnetic field vary on a timescale much larger than $\tau_\eta$.
Note that on the RHS of the above, we can neglect $\bigO(t-\tau)$ contributions to $\frac{ \dh }{\dh t}\!\mesoAvBr{ \FT{\vec{B}}(\vec{p}, t) }$ and use equation \ref{B: eq: fourier space evolution B multiscale final tau_n zero}.
Similarly, we can expand
\begin{equation}
	\exp{\!\left(- \mesoAv{\eta} k^2 \left(t-\tau \right) \right)} = 1 - \left(t-\tau \right) \mesoAv{\eta} k^2 + \bigO( (t-\tau)^2 )
\end{equation}
We then write equation \ref{B: eq: eta fluctuations multiscale average general expression mu b correlator} as
\begin{align}
	\begin{split}
		\mesoAvBr{ \FT{\mu}(\vec{k}-\vec{p},t) \, \FT{\vec{b}}(\vec{p},t) }
		= \FT{Q}\!\left(\vec{k}-\vec{p}\right)
		\vec{p} \cross \left[ \left( \frac{\vec{k} + \vec{p}}{2} \right) \cross \vec{\mathcal{B}}(\vec{k},t) \right] 
	\end{split} \label{eq: <mub> in terms of QmcalB}
\end{align}
where 
\begin{equation}
	\vec{\mathcal{B}}(\vec{k},t) \defn 
	\frac{1}{2}\mesoAvBr{ \FT{\vec{B}}(\vec{k}, t) }
	- \frac{\tau_\eta}{2} \frac{\dh}{\dh t}\!\mesoAvBr{ \FT{\vec{B}}(\vec{k}, t) }
	- \mesoAvBr{ \FT{\vec{B}}(\vec{k}, t) } \frac{\tau_\eta}{2} \, \mesoAv{\eta} k^2
\end{equation}
Using equation \ref{B: eq: fourier space evolution B multiscale final tau_n zero}, we write
\begin{align}
	\begin{split}
		\vec{\mathcal{B}}(\vec{k},t) 
		={}&
		\frac{1}{2}\mesoAvBr{ \FT{\vec{B}}(\vec{k}, t) }
		- \frac{\tau_\eta}{2} \, \evCor(\vec{k}) \mesoAvBr{ \FT{\vec{B}}(\vec{k},t) } 
		- \frac{i \tau_\eta}{2} \, \odCor(\vec{k}) \mesoAvBr{ \FT{\vec{B}}(\vec{k},t) } 
	\end{split} \label{eq: mcalB in terms of meanB}
\end{align}
We note that $\vec{k}\cdot\vec{\mathcal{B}}(\vec{k},t) = 0$ (since $\dive\vec{B} = 0$) and use equations \ref{eq: <mub> in terms of QmcalB} and \ref{eq: mcalB in terms of meanB} to write
\begin{multline}
	\vec{k} \cross \int \d\vec{p} \left(\frac{\vec{k}+\vec{p}}{2}\right)\cross \mesoAvBr{ \FT{\mu}(\vec{k}-\vec{p},t) \, \FT{\vec{b}}(\vec{p},t) }
	\\
	=
	\left[ \evCor(\vec{k}) + i \odCor(\vec{k}) \right]
	\left[ 1 - \tau_\eta \evCor(\vec{k}) - i \tau_\eta \odCor(\vec{k}) \right] \mesoAvBr{ \FT{\vec{B}}(\vec{k},t) }
\end{multline}
where $\odCor$ and $\evCor$ are defined in equations \ref{eq: evCor OdCor definition}.
Putting this in equation \ref{B: eq: mean B multiscale varying eta} and using equation \ref{B.etaFluc: eq: simp cross cross integral meanB homo mean eta}, we write
\begin{align}
	\begin{split}
		\frac{\dh}{\dh t} \mesoAvBr{ \FT{\vec{B}}(\vec{k}) } 
		={}&
		\left[ \evCor(\vec{k}) + i \odCor(\vec{k}) \right] \left[ 1 - \tau_\eta \evCor(\vec{k}) - i \tau_\eta \odCor(\vec{k}) \right] \mesoAvBr{ \FT{\vec{B}}(\vec{k}) }
		- \mesoAv{\eta} k^2 \mesoAvBr{ \FT{\vec{B}}(\vec{k}) } 
		+ \bigO(\tau_\eta^2)
	\end{split} \label{B: eq: B: eq: fourier space evolution B multiscale final first order in tau_n}
\end{align}

\subsection{Growth rate of the large-scale magnetic field}
\label{section: general growth rate}
Let us now focus on the problem of whether a particular Fourier mode of the large-scale magnetic field grows or decays.
We assume $ \mesoAvBr{ \FT{\vec{B}}(\vec{k},t) } \propto \exp\!\left(\lambda t\right) $.
Plugging this into equation \ref{B: eq: B: eq: fourier space evolution B multiscale final first order in tau_n} and taking its real part, we find
\begin{align}
	\begin{split}
		\re(\lambda)
		={}& 
		- \mesoAv{\eta} k^2 
		+ \evCor(\vec{k}) 
		- \tau_\eta \left[ \evCor(\vec{k}) \right]^2
		+ \tau_\eta \left[ \odCor(\vec{k}) \right]^2
		+ \bigO(\tau_\eta^2)
		\label{B: eq: growth rate fluctuating eta multiscale general Q}
	\end{split}
\end{align}
where $\odCor$ and $\evCor$ are defined in equations \ref{eq: evCor OdCor definition}.
From the fact that above, only $\left[ \evCor(\vec{k}) \right]^2$ contains a $k^8$ term, we can see that the growth rate always becomes negative for large-enough $k$ (small-enough scales) as long as $\tau_\eta \ne 0$.
Note that while in the white-noise case, $\odCor(\vec{k})$ only contributed a drift term, it now affects the growth rate as well.

Since we assumed the large-scale magnetic field varies on timescales much larger than $\tau_\eta$, our derivation is self-consistent only when $\abs{\tau_\eta \lambda} \ll 1$.

\section{Dynamo numbers when the fluctuations are isotropic}
\label{section: dyn nums isotrop fluc}
If $Q(\vec{\xi})$ is isotropic, i.e.\@ $Q(\vec{\xi}) = Q(\abs{\vec{\xi}})$ \citep[see][sec.~12.1]{MoninYaglomVol2},
we can write the quantities defined in equation \ref{B.etaFluc: A defn} as
\begin{equation}
	A^{(1)}_i = 0
	\,,\quad
	A^{(2)}_{ij} = \delta_{ij} \, \frac{ A^{(2)}_{mm} }{3}
	\,,\quad
	A^{(3)}_i = 0
\end{equation}
so that
\begin{equation}
	\odCor(\vec{k})
	=
	0
	\,,\quad
	\evCor(\vec{k})
	=
	\frac{ 4 A^{(0)} k^4 - 3 k^2 A^{(2)}_{mm} }{8}
\end{equation}
Equation \ref{B: eq: growth rate fluctuating eta multiscale general Q} can then be written as
\begin{align}
	\begin{split}
		\re(\lambda)
		={}&
		- k^2 \left( \mesoAv{\eta} + \frac{3 A^{(2)}_{mm}}{8} \right)
		+ k^4 \left(\frac{A^{(0)}}{2} - \frac{9 \tau_\eta }{64} \left[ A^{(2)}_{mm}\right]^2 \right)
		\\& + \frac{3 \tau_\eta A^{(0)} A^{(2)}_{mm} k^6 }{8}
		- \frac{\tau_\eta \left[ A^{(0)} \right]^2 k^8 }{4}
		\label{B: eq: growth rate fluctuating eta multiscale general Q assume isotropic}
	\end{split}
\end{align}
If we further define\footnote{
If the correlation function attains a maximum at zero separation, $A^{(2)}_{mm} < 0$.
This implies $\dynOne > 0$.
}
\begin{equation}
	\dynOne \defn
	- \frac{3 A^{(2)}_{mm}}{8 \mesoAv{\eta} }
	\,,\,
	\dynTwo \defn
	\frac{9 \tau_\eta }{32} \, \frac{ \left[ A^{(2)}_{mm}\right]^2 }{ A^{(0)} }
	\label{B.etaFluc.iso: eq: Dyn defn}
	\,,\,
	l_c \defn \sqrt{ - \frac{ 3 A^{(0)} }{ A^{(2)}_{ii} } }
	\,,\,
	K \defn k l_c
	\,,\,
	T \defn \frac{l_c^2}{\mesoAv{\eta} }
\end{equation}
we can write equation \ref{B: eq: growth rate fluctuating eta multiscale general Q assume isotropic} as
\begin{equation}
	T \re(\lambda)
	=
	- K^2 \left( 1 -  \dynOne \right)
	+ \frac{ 4 \dynOne K^4 }{ 9 } \left( 1 - \dynTwo \right)
	- \frac{ 32 \dynOne \dynTwo K^6 }{ 81 }
	- \frac{ 64 \dynOne \dynTwo K^8 }{ 729 }
	\label{B: eq: growth rate fluctuating eta multiscale Q isotropic, scaled vars}
\end{equation}
The growth rate at a particular wavenumber is thus determined by two dynamo numbers, $\dynOne$ and $\dynTwo$.
In appendix \ref{appendix: example gauss correlation function}, we express the dynamo numbers in terms of more observationally relevant quantities by assuming a particular form for the correlation function $Q$.
We see that $\dynOne$ describes how strong the fluctuations of $\eta$ are as compared to its mean value, while $\dynTwo/\dynOne$ is proportional to the ratio of the correlation time of the fluctuations to the diffusion timescale determined from the mean of $\eta$ and the correlation length of its fluctuations.

\begin{figure}
	\centering
	\includegraphics[width=0.6\textwidth,keepaspectratio=true]{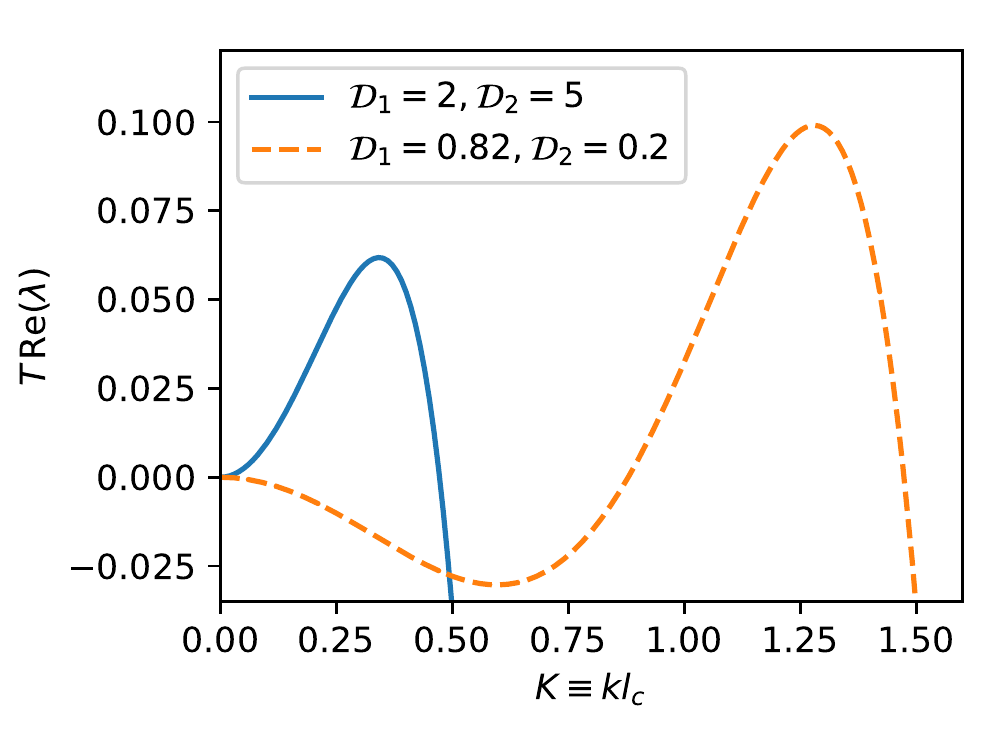}
	\caption{
	The mode growth rate ($T\re(\lambda)$, equation \ref{B: eq: growth rate fluctuating eta multiscale Q isotropic, scaled vars}) as a function of the wavenumber for two combinations of $\dynOne$ and $\dynTwo$.
	}
	\label{B: fig: dynamo due to eta fluctutations}
\end{figure}

Figure \ref{B: fig: dynamo due to eta fluctutations} shows the growth rate (equation \ref{B: eq: growth rate fluctuating eta multiscale Q isotropic, scaled vars}) for two sets of dynamo numbers.
We see that depending on the parameters, the growth rate may peak at large scales or at small scales.

To understand the qualitative behaviour of equation \ref{B: eq: growth rate fluctuating eta multiscale Q isotropic, scaled vars}, we can schematically write it as
\begin{equation}
	\re(\lambda)
	=
	\begin{dcases}
		- k^2 - k^4 - k^6 - k^8 &;\, \dynTwo > 1,\, \dynOne < 1
		\\
		- k^2 + k^4 - k^6 - k^8 &;\, \dynTwo < 1,\, \dynOne < 1
		\\
		k^2 + k^4 - k^6 - k^8 &;\, \dynTwo < 1,\, \dynOne > 1
		\\
		k^2 - k^4 - k^6 - k^8 &;\, \dynTwo > 1,\, \dynOne > 1
	\end{dcases}
\end{equation}
In the first case, $\re(\lambda)$ is always negative, and so there is no dynamo.
In the last two cases, $\re(\lambda)$ is positive for small $k$ and becomes negative for large wavenumbers.
In the second regime, it seems to be difficult to say anything concrete (depending on the values of the coefficients, one can either have growth in a range of wavenumbers or growth nowhere).

\begin{figure}
	\centering
	\includegraphics[width=0.48\textwidth,keepaspectratio=true]{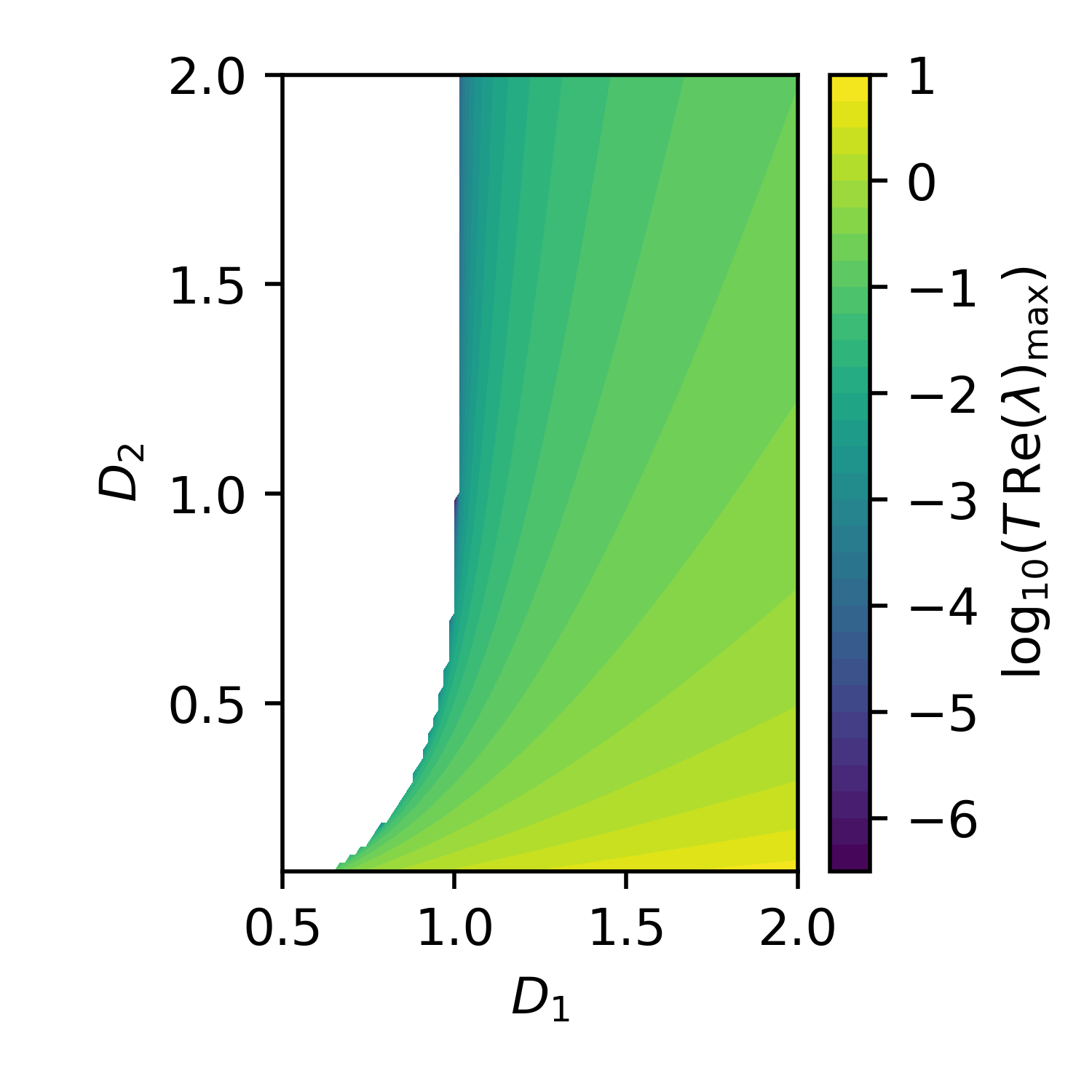} %
	\includegraphics[width=0.51\textwidth,keepaspectratio=true]{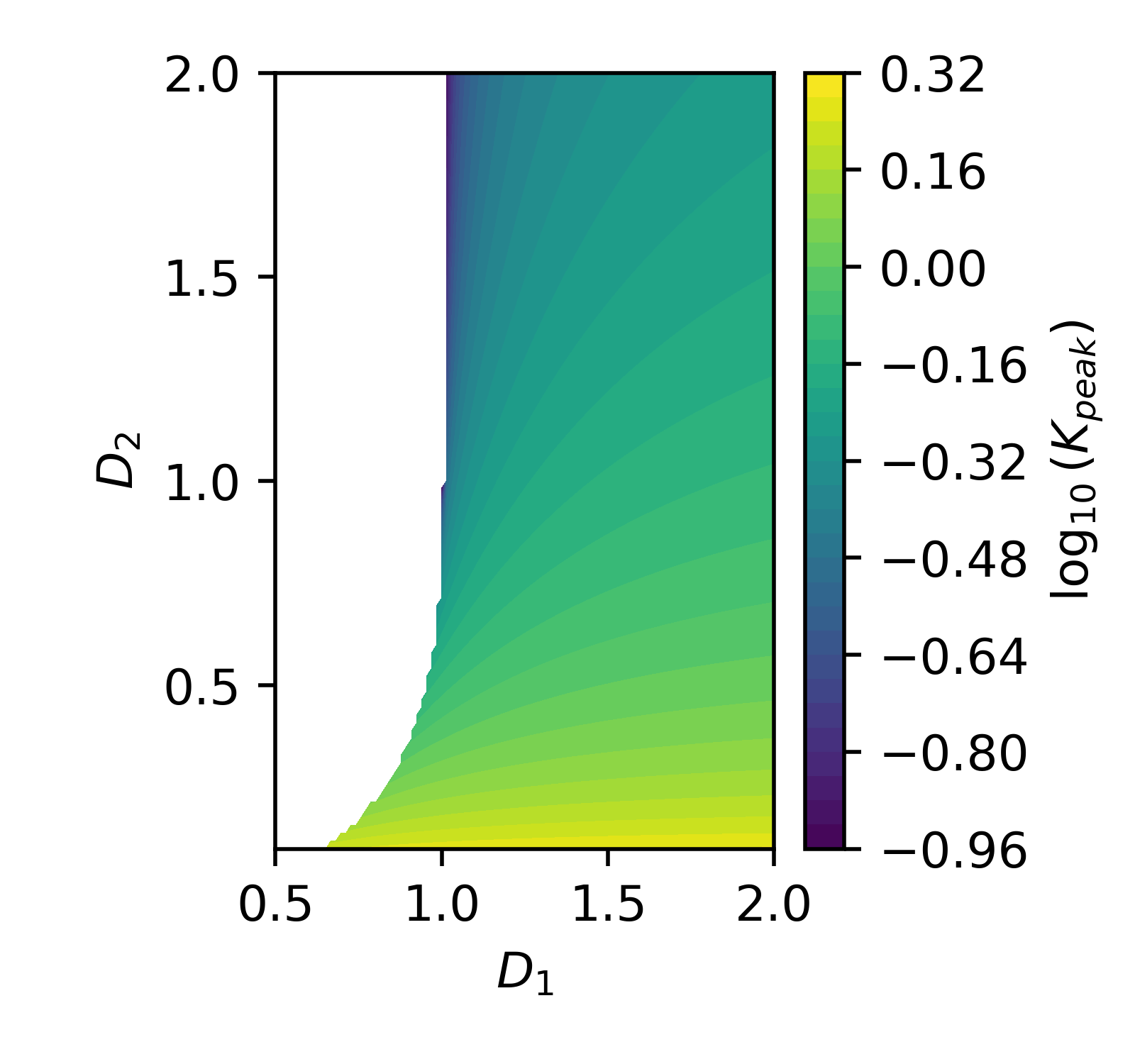}
	\caption{
	Left: 
	The peak growth rate ($T\re(\lambda)$, equation \ref{B: eq: growth rate fluctuating eta multiscale Q isotropic, scaled vars}).
	In the white regions, the growth rate is negative for all $K$.
	Right:
	The wavenumber ($K$) at which the growth rate peaks.
	Recall that for a mode with wavelength $l_c$, the wavenumber would be $K=2\pi \approx 10^{0.8}$.
	}
	\label{fig: dynamo iso countour gr and Kpeak}
\end{figure}

Since \ref{B: eq: growth rate fluctuating eta multiscale Q isotropic, scaled vars} is a polynomial in $K$, one can easily solve for its extrema.
In figure \ref{fig: dynamo iso countour gr and Kpeak}, we show the dynamo growth rate (where positive) and the wavenumber of the resulting large-scale field, as a function of $\dynOne$ and $\dynTwo$.

If we drop the terms of order $K^6$ and $K^8$ in equation \ref{B: eq: growth rate fluctuating eta multiscale Q isotropic, scaled vars} (this does not change the qualitative behaviour when $\dynTwo > 1$), we can estimate that if $\dynOne > 1$, the growth rate attains a maximum value at $K_\text{peak}$, where
\begin{equation}
	K_\text{peak} \approx \sqrt{ \frac{9 \left(\dynOne - 1\right)}{8 \dynOne \left(\dynTwo-1\right)} }
	\,,\quad
	\left[T \re(\lambda) \right]_\text{max} \approx \frac{9 \left(\dynOne - 1\right)^{2}}{16 \dynOne \left(\dynTwo - 1\right)}
\end{equation}

Broadly speaking, there are two kinds of regimes in which the dynamo is excited.
One, $\dynOne > 1$, corresponds to the fluctuations being strong enough that the effective diffusivity itself becomes negative (but the growth itself is still cut off at small scales due to higher-order terms).
The other, $\dynTwo < 1$ (with $\dynOne$ also $< 1$), corresponds to growth with the effective diffusion remaining positive;
one can see, however, from figure \ref{fig: dynamo iso countour gr and Kpeak} that this growth happens at smaller scales than in the other regime (but may still be at scales larger than $l_c$).
While $\dynTwo \ll 1$ can formally lead to growing solutions regardless of the value of $\dynOne$, the growth then occurs at scales $\lesssim l_c$.

\section{The effect of anisotropy}
\label{B.etaFluc: section: anisotropy}
Although we have not done so in the above, it seems natural to assume that the temporal correlation function $S$, that appears in equation \ref{B: eq: homogeneous separable mumu at large scale, fourier space}, is even.
This would allow one to take $\int_{-\infty}^\infty S(t) \, \d t = 1$ and define the correlation time of $\eta$ as $\tau_\eta \defn \int_{-\infty}^\infty \abs{t} S(t) \, \d t$.

Because $\mu$ is a scalar, assuming its double correlation is invariant under time-reversal immediately implies $\FT{Q}(\vec{k}) = \FT{Q}(-\vec{k})$.
We then conclude that
\begin{equation}
	A^{(1)}_i = A^{(3)}_i = \odCor(\vec{k}) = 0
\end{equation}
when the fluctuations of $\eta$ are separable, homogeneous, stationary, and time-reversal-invariant; this holds even without assuming that the fluctuations of $\eta$ are isotropic!
We now study the dynamo assuming the double correlation of $\mu$ is time-reversal invariant and anisotropic.

Let us choose the coordinate axes $1,2,3$ to be along the principal axes of the matrix $A^{(2)}$ (defined in equation \ref{B.etaFluc: A defn}), with the corresponding eigenvalues being $-a_1$, $-a_2$, and $-a_3$ (such that $a_1 \ge a_3 \ge a_2$).
By analogy with equation \ref{B.etaFluc.iso: eq: Dyn defn}, one can define the correlation length along each axis as $l_c^{(i)} \defn \sqrt{A^{(0)}/a_i}$.
It is physically reasonable to assume $Q(\vec{\xi})$ attains a local maximum at the origin, and that its correlation length is finite.
This means $a_1, a_2, a_3 > 0$.

Analogous to equation \ref{B.etaFluc.iso: eq: Dyn defn}, we define
\begin{equation}
	\dynOne \defn
	\frac{9 a_3}{8 \mesoAv{\eta} }
	\,,\,
	\dynTwo \defn
	\frac{81 \tau_\eta a_3^2 }{32 A^{(0)} }
	\,,\,
	l_c \defn \sqrt{ \frac{ A^{(0)} }{ a_3 } }
	\,,\,
	\vec{K} \defn \vec{k} l_c
	\,,\,
	T \defn \frac{l_c^2}{\mesoAv{\eta} }
\end{equation}
We also define the new quantities
\begin{equation}
		\arOne
		\defn
		\frac{a_1}{a_3} - 1 
	\,,\,
		\arTwo
		\defn
		1 - \frac{a_2}{a_3}
	\,,\,
		n_1 \defn \frac{ \abs{K_1} }{K}
	\,,\,
		n_2 \defn \frac{ \abs{K_2} }{K}
\end{equation}
and a modified dynamo number
\begin{equation}
	\dynOneA
	\defn
	\dynOne\left[ 
		1
		+ \frac{\arOne}{9} \left( 2 + 3 n_1^2 \right)
		- \frac{\arTwo}{9} \left( 2 + 3 n_2^2 \right)
	\right]
\end{equation}
Since $0 \le \arTwo < 1$, $0 \le \arOne < \infty$, and $0 \le n_1, n_2 \le 1$, we find that $\dynOneA > 0$.
We write  $\evCor(\vec{k})$ (equation \ref{eq: evCor OdCor definition}) as
\begin{align}
	\begin{split}
		T \evCor(\vec{k})
		={}&
		\dynOneA K^2
		+ \frac{4\dynOne}{9} \, K^4
		\label{B.etaFluc.aniso: eq: Tg in terms of dynamo numbers}
	\end{split}
\end{align}
Noting that $\tau_\eta/T = 4\dynTwo/(9\dynOne)$, one can substitute equation \ref{B.etaFluc.aniso: eq: Tg in terms of dynamo numbers} in equation \ref{B: eq: growth rate fluctuating eta multiscale general Q} to obtain the following expression for the growth rate:
\begin{equation}
	T \re(\lambda)
	=
	K^{2} \left(\dynOneA - 1\right)
	+ \frac{ 4\dynOne K^{4} }{9} \left(1 - \frac{\dynTwo \dynOneA^{2} }{ \dynOne^2 }\right)
	- \frac{32 \dynTwo \dynOneA K^{6}}{81}
	- \frac{64 \dynOne \dynTwo K^{8}}{729}
	\label{B: eq: growth rate fluctuating eta multiscale Q aniso, scaled vars}
\end{equation}
\begin{figure}
	\centering
	\includegraphics[width=0.6\textwidth,keepaspectratio=true]{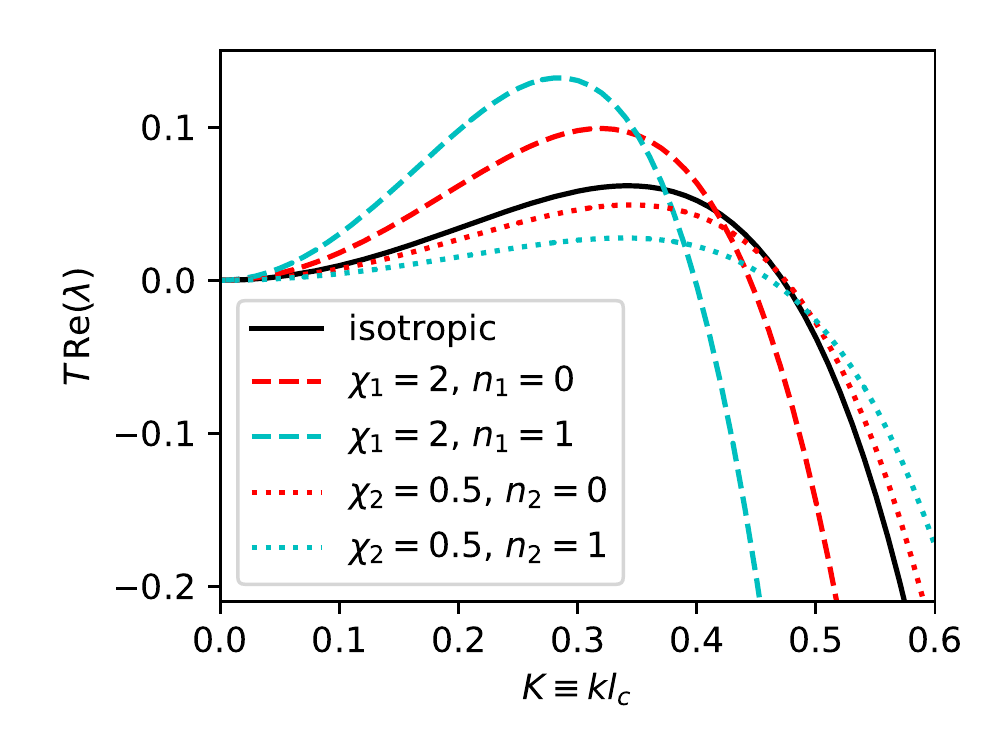}
	\caption{
	The mode growth rate ($T\re(\lambda)$, equation \ref{B: eq: growth rate fluctuating eta multiscale Q aniso, scaled vars}) as a function of the wavenumber.
	In all the cases, we have taken $\dynOne = 2$ and $\dynTwo = 5$.
	Other parameters not mentioned in the legend have been set to zero.
	}
	\label{B: fig: dynamo due to eta fluctutations aniso}
\end{figure}
As expected, this reduces to equation \ref{B: eq: growth rate fluctuating eta multiscale Q isotropic, scaled vars} on setting $\dynOneA = \dynOne$.
Replacing $\dynOne \to \dynOneA$ and $\dynTwo \to \dynTwo \dynOneA^{2}/\dynOne^2$, our comments in section \ref{section: dyn nums isotrop fluc} on the qualitative behaviour of equation \ref{B: eq: growth rate fluctuating eta multiscale Q isotropic, scaled vars} also apply to this equation.
Unlike in the isotropic case, the growth rate now depends on the direction of $\vec{K}$ through the direction cosines $n_1$ and $n_2$.
Figure \ref{B: fig: dynamo due to eta fluctutations aniso} shows the growth rate as a function of the wavenumber for various parameter values.

\section{Suppression of turbulent diffusion}
\label{section: general suppression turb diff}

Neglecting terms with more than two spatial derivatives of $\mesoAvBr{ \vec{B} } $, equation \ref{B: eq: B: eq: fourier space evolution B multiscale final first order in tau_n} can be written as
\begin{align}
	\begin{split} 
		\frac{\dh}{\dh t} \mesoAvBr{ \FT{\vec{B}}(\vec{k}) } 
		={}&
		- \mesoAv{\eta} k^2 \mesoAvBr{ \FT{\vec{B}}(\vec{k}) }
		- \frac{1}{8} \left( 3 k_mk_n A^{(2)}_{mn} + 2 k^2 A^{(2)}_{mm} \right) \mesoAvBr{ \FT{\vec{B}}(\vec{k}) }
		\\& + \frac{i}{8} \, k_m A^{(3)}_m \mesoAvBr{ \FT{\vec{B}}(\vec{k}) }
		+ \frac{\tau_\eta }{64} \, k_m k_i A^{(3)}_m A^{(3)}_i \mesoAvBr{ \FT{\vec{B}}(\vec{k}) }
		+ \bigO(k^3)
	\end{split} \label{B: eq: fourier space evolution B multiscale final first order in tau_n neglect k^3}
\end{align}
Following the reasoning used in section \ref{B.etaFluc: section: anisotropy} for $\dynOneA$, one can see that the coefficient of $\mesoAvBr{\FT{\vec{B}}(\vec{k}) }$ in the second term above is always positive as long as the spatial correlation function of the $\eta$-fluctuations attains a maximum at zero separation; the turbulent diffusion is suppressed.
This can be seen more clearly in equation \ref{B:.etaFluc: white noise evol assume simple corrfun}, which assumes a particular form for the spatial correlation function.
The third term just describes advection with an effective velocity $\vec{A}^{(3)}/8$,
analogous to `Moffatt drift' \citep[sec.~7.11]{MoffatMagFieldGenBook}.
The fourth term can never be negative, and is nonzero only when the fluctuations of $\eta$ are anisotropic and not invariant under time-reversal.
As noted in section \ref{section: dyn nums isotrop fluc}, the higher powers of $k$ neglected in equation \ref{B: eq: fourier space evolution B multiscale final first order in tau_n neglect k^3} can cause growth of the large-scale magnetic field even when the effective diffusivity is positive.
They also ensure that the growth rate becomes negative at small scales.

It may seem counter-intuitive that a dissipative term ($\eta$) at the mesoscale leads to a dynamo at larger scales, but it must be noted that in addition to dissipation, $\eta$ also contributes an effective advection term (usually referred to as `turbulent diamagnetism'; see equation \ref{B: eq: diamagnetic pumping}) when
spatial variations at the mesoscale are properly accounted for.
Heuristically, it seems possible to explain the suppression of turbulent diffusion by turbulent diamagnetism as follows:
turbulent diamagnetism causes the magnetic field to be preferentially concentrated in regions where the turbulent diffusivity is minimal.
The effective turbulent diffusivity acting on the magnetic field is then less than what would be inferred by taking an average over the entire system.
See \citet[p.~49]{silantev99} for a more general explanation of reduced turbulent diffusion when two scattering mechanisms contribute to the diffusion process.

\section{Estimates of the dynamo numbers}
\label{B.etaFluc: estimate dynamo numbers}
Unfortunately, fluctuations of the turbulent diffusivity in astrophysical systems are not sufficiently constrained by observations.
The situation in the solar photosphere is comparatively better, as observations of granulation give us an idea of the order of magnitude of various quantities.
To make crude estimates, we use equation \ref{eq: D1 D2 for gauss corr} which assumes a specific form for the correlation function of $\eta$.

Let us assume $l_c = \SI{3}{Mm}$ \citep[peak of the granulation's power spectrum as observed by][fig.~2]{RoMu86} and $\tau_\eta = \SI{400}{s}$ \citep[granule lifetime measured by][]{BaSc61}.
The turbulent diffusivity in the photosphere is a scale-dependent quantity, which is moreover not very well constrained \citep[fig.~10]{abramenko11}.
For the length scales of interest, it is not unreasonable to take $\mesoAv{\eta} = \SI{600}{km^2.s^{-1}}$.
Let us also assume  $f=0.1$ ($f \defn \mesoAvBr{\mu^2}/\mesoAv{\eta}^2$).
We then find $\dynOne \approx \num{6e-3}$ and $\dynTwo \approx \num{4e-4}$.
These estimates appear to rule out the operation of such a dynamo in the solar photosphere.
However, we note that assuming slightly different values of $l_c$ and $\tau_\eta$ brings the dynamo numbers to within the regime where a large-scale field can be generated;
for example, taking $l_c = \SI{300}{km}$ and $\tau_\eta = \SI{900}{s}$ gives us $\dynOne \approx 1.4$ and $\dynTwo \approx 18$.
The dynamo numbers are also affected by uncertainties in $f$.
Further, anisotropy can have a significant effect on the growth rates.
Better estimates of the dynamo numbers would require measurements of the spatiotemporal correlation and strength of fluctuations of the turbulent diffusivity (or the kinetic energy) in the solar photosphere.

\section{Conclusions}
\label{section: conclusions}

We have used a double-averaging procedure and found that just like helicity fluctuations, fluctuations of the turbulent kinetic energy can drive the growth of a large-scale magnetic field.
While \citet[p.~49]{silantev99} has also reported that spatiotemporal fluctuations of the turbulent kinetic energy reduce the effective turbulent diffusion, we are not aware of any detailed studies of this effect that consistently account for the concomitant spatial gradients.

In the white-noise limit, we have found that $\eta$-fluctuations cause a reduction in the overall turbulent diffusion (in agreement with previous work), while also contributing a drift term which does not affect the growth of the field.
We have then explored effects of nonzero correlation times and found the possibility of growing mean field solutions despite the overall turbulent diffusion remaining positive.
When the fluctuations are isotropic, the growth rate 
of a particular Fourier mode of the large-scale magnetic field depends on the magnitude of its wavevector and on two dynamo numbers.
Anisotropy leads to a dependence on, among other things, the direction of the wavevector.

We have studied the conditions under which this new dynamo can operate.
However, the lack of precise estimates of the quantities involved makes it hard to conclusively rule out or support the resulting dynamo in various astrophysical scenarios.

Given the prevalence of shear in astrophysical systems, an obvious extension of the current work would be to study the implications, for a large-scale magnetic field, of fluctuations of the turbulent kinetic energy in a shearing background.
Since inhomogeneities in the density and in the small-scale magnetic energy also give rise to pumping \citep{vainshtein1983macroscopic}, we expect them to have effects similar to those described here.

\backsection[Acknowledgements]{
We thank the anonymous referees for useful comments.
We thank Alexandra Elbakyan for facilitating access to scientific literature.
}

\backsection[Funding]{
This research received no specific grant from any funding agency, commercial or not-for-profit sectors.
}

\backsection[Software]{
Numpy \citep{numpy2020} and Matplotlib \citep{matplotlib2007}.
}

\backsection[Declaration of interests]{
The authors report no conflict of interest.
}

\backsection[Author ORCID]{
KG, \url{https://orcid.org/0000-0003-2620-790X}; 
NS, \url{https://orcid.org/0000-0001-6097-688X}
}

\backsection[Author contributions]{
KG and NS conceptualized the research, interpreted the results, and wrote the paper.
KG performed the calculations.
}

\bibliographystyle{jfm}
\bibliography{refs.bib}

\appendix
\section{Dynamo numbers for a simple correlation function}
\label{appendix: example gauss correlation function}
To physically interpret $\dynOne$ and $\dynTwo$ (defined in equation \ref{B.etaFluc.iso: eq: Dyn defn}), it is helpful to explicitly write them out for a specific functional form of $Q$ (see equation \ref{B: eq: homogeneous separable mumu at large scale, fourier space}).
We take
\begin{equation}
	Q(\vec{\xi}) = C \exp\!\left( - \frac{ \abs{\vec{\xi}}^2 }{2 l_c^2}\right)
	\,, \quad
	S(t) = \frac{1}{2\tau_\eta} \exp\!\left(- \frac{\abs{t}}{\tau_\eta} \right)
	\label{B.etaFluc: eq: correlation functions}
\end{equation}
which gives us
\begin{equation}
	A^{(0)} = C > 0
	\,,\quad
	A^{(1)}_i = 0
	\,,\quad
	A^{(2)}_{ij} = - \frac{C}{l_c^2 } \, \delta_{ij}
	\,,\quad
	A^{(3)}_{i} = 0
\end{equation}
If we define
\begin{equation}
	\tilde{\tau} \defn \frac{\tau_\eta}{T} = \frac{\tau_\eta \mesoAv{\eta} }{l_c^2}
	\,,\quad
	f \defn \frac{ \mesoAvBr{\mu^2} }{ \mesoAv{\eta}^2 }
\end{equation}
and use the fact that $ \mesoAvBr{\mu^2} = C/(2\tau_\eta) $ (recall that $\mu \defn \eta - \mesoAv{\eta}$), the dynamo numbers (equation \ref{B.etaFluc.iso: eq: Dyn defn}) become
\begin{equation}
	\dynOne \defn \frac{9 f\tilde{\tau} }{4} 
	\,,\quad
	\dynTwo \defn \frac{81 f \tilde{\tau}^2 }{16} 
	\label{eq: D1 D2 for gauss corr}
\end{equation}
Note that when $\tilde{\tau}\to 0$, $\dynOne$ remains constant, while $\dynTwo\to 0$.
Here, $f$ represents the strength of the fluctuations of $\eta$, while $\tilde{\tau}$ is a scaled measure of their correlation time.

\section{What if we did not have turbulent diamagnetism?}
\label{appendix: no pumping}
Instead of equation \ref{B: eq: diamagnetic pumping}, let us consider the following expression for the EMF:
\begin{equation}
	\vec{\emf} = - \eta \curl\vec{B}
	\label{B.etaFluc.nodia: eq: diamagnetic pumping}
\end{equation}
Equation \ref{B: eq: mean B multiscale varying eta} is then replaced by
\begin{equation}
	\frac{\dh}{\dh t} \mesoAvBr{ \FT{\vec{B}}(\vec{k}) }
	=
	\int\d\vec{p}\, \vec{k} \cross \left[ \vec{p} \cross \left\{ \mesoAvBr{ \FT{\eta}(\vec{k}-\vec{p}) }\mesoAvBr{ \FT{\vec{B}}(\vec{p}) } + \mesoAvBr{ \FT{\mu}(\vec{k}-\vec{p}) \,\FT{\vec{b}}(\vec{p}) } \right\} \right]
	\label{B.etaFluc.nodia: eq: mean B multiscale varying eta}
\end{equation}
Equation \ref{B: eq: dbBYdt fourier space} is replaced by
\begin{align}
	\begin{split}
		\frac{\dh \FT{\vec{b}}(\vec{k}) }{\dh t}
		={}&
		\int\d\vec{p}\, \vec{k} \cross \bigg[ \vec{p} \cross \bigg( 
		\mesoAvBr{ \FT{\eta}(\vec{k}-\vec{p}) } \FT{\vec{b}}(\vec{p})
		+ \FT{\mu}(\vec{k}-\vec{p}) \mesoAvBr{ \FT{\vec{B}}(\vec{p}) }
		\bigg)  
		\bigg]
	\end{split} \label{B.etaFluc.nodia: eq: dbBYdt fourier space}
\end{align}
Equations \ref{B.etaFluc: eq: simp cross cross integral homo mean eta} and \ref{B.etaFluc: eq: simp cross cross integral meanB homo mean eta} remain unchanged.
Equation \ref{B: eq: eta fluctuations multiscale average general expression mu b correlator} becomes
\begin{multline}
	\mesoAvBr{ \FT{\mu}(\vec{q},t) \FT{\vec{b}}(\vec{k},t) }
	\\=
	\int_{-\infty}^t \d\tau \,
	e^{- \mesoAvBr{\eta} k^2 \left(t-\tau \right)}  
	\, \vec{k} \cross \left[ \left( \vec{k} + \vec{q} \right) \cross \mesoAvBr{ \FT{\vec{B}}(\vec{k}+\vec{q}, \tau) } \right] 
	\FT{Q}\!\left( \vec{q} \right) S(t-\tau)
	\label{B.etaFluc.nodia: eq: eta fluctuations multiscale average general expression mu b correlator}
\end{multline}
Assuming $S(t) = \Dirac(t)$ and plugging equation \ref{B.etaFluc.nodia: eq: eta fluctuations multiscale average general expression mu b correlator} into equation \ref{B.etaFluc.nodia: eq: mean B multiscale varying eta}, we obtain the following evolution equation for the large-scale magnetic field:
\begin{align}
	\begin{split}
		\frac{\dh}{\dh t} \mesoAvBr{ \FT{\vec{B}}(\vec{k}) }
		={}&
		- \mesoAv{\eta} k^2 \mesoAvBr{ \FT{\vec{B}}(\vec{k}) }
		+ \frac{1}{2} \int\d\vec{s} \, \FT{Q}\!\left( \vec{s} \right) \left( k^2 \delta_{ij} - k_i k_j \right) s_j \vec{s} \cdot \mesoAvBr{ \FT{\vec{B}}(\vec{k}) }
		\\& + \frac{1}{2} \int\d\vec{s} \, \FT{Q}\!\left( \vec{s} \right) \left( k^2 - \vec{k} \cdot \vec{s} \right)^2 \mesoAvBr{ \FT{\vec{B}}(\vec{k}) }
	\end{split}
\end{align}

The second term on the RHS above is qualitatively different from any term present in equation \ref{B: eq: fourier space evolution B multiscale final tau_n zero}; due to this term, the various components of the large-scale magnetic field may become coupled when the fluctuations of $\eta$ are anisotropic.
This equation (or its extension to the case of nonzero correlation time) may also be used to describe scenarios where the microscopic conductivity itself exhibits stochastic fluctuations.
\citet{PetAleGis16} and \citet{GreRudEls23} have studied such systems.

\end{document}